\documentclass[pre,twocolumn,showpacs]{revtex4}
\usepackage{xcolor}
\usepackage{framed}
\definecolor{shadecolor}{rgb}{1,1,0}
\usepackage{color,soul}
\usepackage{amsmath}
\usepackage{amssymb}
\usepackage{graphicx}
\usepackage{array}
\usepackage{setspace}
\usepackage[sort&compress]{natbib}
\bibpunct{[}{]}{,}{}{}{}

\usepackage{graphicx}
\usepackage{amsmath}
\usepackage{amsfonts}
\def\be{\begin{equation}}
\def\ba{\begin{eqnarray}}

\def\b{\beta}

\def\d{\delta}
\def\D{\Delta}
\def\e{\epsilon}

\def\Th{\Theta}

\def\l{\lambda}

\def\p{\pi}
\def\r{\rho}

\def\f{\varphi}

\def\ps{\psi}
\def\o{\omega}
\def\O{\Omega}
\def\i{\int}
\def\Tr{\mbox{Tr}}
\def\bz{{\mathbf z}}
\def\ee#1{\label{#1}\end{equation}}
\def\ea#1{\label{#1}\end{eqnarray}}

\begin{document}
\title{Microcanonical quantum fluctuation theorems}
\author{Peter Talkner,
Peter H\"anggi}
\affiliation{Institute of Physics, University of Augsburg,
D-86135 Augsburg, Germany}
\author{Manuel Morillo}
\affiliation{Fisica Te\'orica, Universidad de Sevilla, Apartado de Correos
  1065, Sevilla 41080, Spain}
\date{\today}
\begin{abstract}
Previously derived expressions for the characteristic function of work
performed on a quantum system  by a classical external force are
generalized to arbitrary initial states of the considered system and
to Hamiltonians with degenerate spectra. In the particular case of
microcanonical initial states explicit expressions for the
characteristic function and the corresponding probability density of
work are formulated. Their classical limit as well as their relations
to the respective canonical expressions are discussed. A fluctuation
theorem is derived that expresses the ratio of probabilities of work for a
process and its time reversal 
to the  ratio of densities of states of
the microcanonical equilibrium systems with corresponding initial and final
Hamiltonians.
From this Crooks-type fluctuation theorem a relation between 
entropies of different systems can be derived which does not involve the time
reversed process. This entropy-from-work theorem provides an 
experimentally accessible way to
measure entropies.

\end{abstract}

\maketitle
\section{Introduction}
During the last decade various fluctuation and work theorems
\cite{ECM,GC,J,C} have been formulated and discussed. They provide
information about the fully {\it nonlinear} response of a system
under the action of a time dependent force \cite{BK,JLNP,SS,J2} 
in contrast to linear response theories in which the response is
expressed in terms of correlation functions of the unperturbed system
\cite{CW,Ku,GHT,Hthesis,HT}.  
Moreover, these
theorems were used to establish the second law of thermodynamics and
to sharpen its formulation \cite{AN,MT,KPvdB}. Fluctuation and work 
theorems have
primarily been derived, numerically tested and experimentally
confirmed for classical systems which initially, or permanently are
in contact with a heat bath \cite{DCPR,BLR,LD}. The system then
initially stays
in a state 
described by a canonical (Maxwell-Boltzmann)
distribution at the temperature of
the heat bath. Classical isolated systems which initially are in a
microcanonical state were investigated in Ref.~\cite{CvdBK}. 
Quantum mechanical generalizations
were proposed recently \cite{K,Tas,Mu,ME,TLH,TH}, but only 
for canonical initial states.
The equivalence of canonical and microcanonical initial states was
demonstrated for classical systems in the {\it thermodynamic limit} \cite{PS}.
Little emphasis has been given to the impact of general initial
conditions on the work performed on a quantum system by a classical
force \cite{ANpre}.

In the present paper we investigate the work performed by an external
force acting on an otherwise isolated quantum system. The characteristic
function of the work is shown to always assume the form of a correlation
function of the
exponentiated system Hamiltonians at the initial and final time,
regardless of how the system was initially prepared.
This
characteristic function comprises all aspects of the statistics of
the work.
In the special case of a microcanonical initial state the inverse
Fourier transform leads to an expression for the density of performed
work which directly leads to a quantum generalization of the
microcanonical version of Crooks theorem. This theorem relates the
probability densities of work of the processes and its time reversed
partner process 
to the difference of entropies of the
equilibrium states that correspond to the initial and final
Hamiltonians. Such relations have been confirmed for classical systems
by computer
experiments \cite{CvdBK}. In real experiments an active time reversal
is not feasible. We yet present an ``entropy-from-work'' theorem that does not
contain the time reversed process and still allows one to obtain the
entropy of the equilibrium system for the system with the final
Hamiltonian, provided the entropy of the system with the initial
Hamiltonian is known. 

Apart from its use in numerical
  investigations such as molecular dynamics \cite{PHT,BHKPV} or
  microcanonical Monte Carlo 
  simulations \cite{MC}, the microcanonical ensemble is known to
  provide the valid  description of isolated systems in equilibrium
  \cite{Callen}. It presents the proper statistical mechanical
    framework for the description of isolated systems of finite size
    such as  clusters of
  atoms \cite{SKHDKIH}, atomic nuclei \cite{BBIMS} 
or Bose Einstein condensates\cite{HKK} and even allows
for phase transitions in finite systems \cite{T,HD,JBJ}.

The paper is organized as follows. In Sect.~\ref{II} the
characteristic function of work is expressed as a correlation
function of exponentiated Hamiltonians. The density matrix which
enters this expression is given by the initial density matrix of the
system projected onto the diagonal elements with respect to the
eigenbasis of the initial Hamiltonian. In the special case of a
canonical density matrix the known form of the canonical
characteristic function is recovered. In Sect.~\ref{III} we consider
a microcanonical initial state, and derive microcanonical quantum 
fluctuation and work theorems. Moreover, the probability densities and
characteristic functions for canonical 
initial states are shown to be related to the respective
microcanonical quantities by properly
weighted Laplace transforms. Conclusions are presented with a final
Sect. IV.

\section{Characteristic functions of work}
\label{II}
The response of a quantum system on the perturbation by a classical,
external force can be characterized by the change of energy
contained in the total system. The energy of
the total system is determined by its Hamiltonian $H(t)$. It includes
the external, time-varying force and therefore depends on time. We
will consider the 
dynamics of the system only within a finite window of time $[t_0,t_f]$
during which the force is acting in a prescribed way, resulting in a
protocol of Hamiltonians, which we denote by $\{H(t)\}_{t_f,t_0}$.
A measurement of the Hamiltonian $H(t)$ at a time $t$ will result in
an energy which is an eigenvalue $e_k(t)$ of the Hamiltonian, i.e.
\be
H(t) \f_{k,\l}(t) = e_k(t) \f_{k,\l}(t),
\ee{Hfe}
where $\l$ is a quantum number which labels a possible degeneracy of
the eigenvalue $e_k(t)$. The eigenfunctions
$\f_{k,\l}(t)$ can be chosen as normalized and pairwise
orthogonal. They span the eigenspace of $H(t)$ belonging to
$e_k(t)$. 
The projection operator on this eigenspace becomes
\be
P_k(t) = \sum_\l |\f_{k,\l}(t) \rangle \langle  \f_{k,\l}(t) |.
\ee{Pk}

Measuring the Hamiltonian  at the
respective times of measurement $t_0$ and $t_f$, one obtains as results
eigenvalues 
$e_n(t_0)$ and $e_m(t_f)$ of the Hamiltonian $H(t)$.  The work $w$
performed on the system is given by the difference of the measured
energies, i.e. by
\be w = e_m(t_f)-e_n(t_0). \ee{w} 
The energy values $e_n$ and $e_m$ arising from the measurements are
  random quantities. Consequently, the observed value of work, $w$, too is a
  random quantity. 
The probability $p_n$, with which the particular
eigenvalue $e_n(t_0)$ is observed in the first measurement, depends
on the density matrix $\r(t_0)$ which describes the state of the
total system at time $t_0$. According to the laws of quantum
mechanics this probability is given by the expectation value of the
projection operator $P_n(t_0)$ onto the subspace of eigenstates of
the Hamiltonian $H(t_0)$ with energy $e_n(t_0)$, i.e. by 
\be 
p_n =
\Tr P_n(t_0) \r(t_0). 
\ee{pn} 
Immediately after this measurement the
system is found in the corresponding state with properly normalized density
matrix 
\be \r_n = \frac{P_n(t_0) \r(t_0) P_n(t_0)}{p_n} 
\ee{rn} 
and
then evolves in time according to 
\be \r_n(t) = U_{t,t_0} \r_n
U^+_{t,t_0}, 
\ee{rtt0} 
where the unitary time evolution operator
$U_{t_f,t_0}$ obeys the Schr\"odinger equation \be i \hbar \partial
U_{t,t_0}/\partial t = H(t) U_{t,t_0}, \quad U_{t_0,t_0}=1. \ee{U}
Note, that in general the  time evolved states $\ps_{n,\l}(t) =
U_{t,t_0} \f_{n,\l}(t_0)$
  are not eigenfunctions of $H(t)$. An exception are quasi-static
    changes of the Hamiltonian for which the adiabatic theorem holds \cite{at}
The second measurement, at time $t_f$, produces an eigenvalue $e_m(t_f)$
of $H(t_f)$ which occurs with probability
\be
\begin{split}
p(m|n) &= \Tr P_m(t_f) \r_n(t_f)\\
 &= \Tr P_m(t_f) U_{t_f,t_0}\r_n U^+_{t_f,t_0}.
\end{split}
\ee{pnm} The probability density of work, $p_{t_f,t_0}(w)$ can then
be expressed as 
\be p_{t_f,t_0}(w) = \sum_{m,n} \d\left (
w-[e_m(t_f) -e_n(t_0)] \right ) p(m|n)p_n. 
\ee{ptft0} 
For the
characteristic function of the work, 
\be G_{t_f,t_0}(u) = \i dw\:
e^{iuw} p_{t_f,t_0}(w), 
\ee{G} 
being the Fourier transform of the
probability density of work, cf. e.g. \cite{HT,vK}, we find 
\be
\begin{split}
G_{t_f,t_0}(u) = &\: \sum_{m,n} e^{iu e_m(t_f)} e^{-iue_n(t_0)} p(m|n) p_n \\
=&\: \sum_{m,n}  e^{iu e_m(t_f)} e^{-iue_n(t_0)}\\
&\:\times \Tr P_m(t_f) U_{t_f,t_0} \r_n
U^+_{t_f,t_0} p_n \\
=&\: \sum_{m,n} \Tr\: e^{iu H(t_f)} P_m(t_f) \\
& \:\times U_{t_f,t_0} e^{-iu H(t_0)} \r_n
U^+_{t_f,t_0}\:
p_n\\
=& \:\Tr \:e^{iu H(t_f)} U_{t_f,t_0} e^{-iu H(t_0)} \bar{\r}(t_0)
U^+_{t_f,t_0}\\
=& \:\Tr \: e^{iu H_{H}(t_{f})} e^{-iu H(t_{0})} \bar{\r}(t_{0})\\
\equiv&\: \langle e^{iu H(t_f)} e^{-iu H(t_0)} \rangle_{t_0},
\end{split}
\ee{Gu}
where 
\be
H_{H}(t_{f}) = U^{+}_{t_{f},t_{0}} H(t_{f})U_{t_{f},t_{0}}
\ee{HH}
denotes the Hamiltonian in the Heisenberg picture.
Further, we used the completeness relation $\sum_m P_m(t_f) = 1$ and
introduced the density matrix
\be
\bar{\r}(t_0) = \sum p_n \r_n  = \sum P_n \r(t_0) P_n
\ee{rb}
with respect to which the average in the correlation function $\langle
e^{iu H(t_f)} e^{-iu H(t_0)} \rangle_{t_0} $ is performed.
This density matrix describes the state of the system
projected onto the eigenbasis of the Hamiltonian at time $t=t_0$. It
coincides with the initial density matrix $\r(t_0)$ only if $\r(t_0)$
commutes with the Hamiltonian $H(t_0)$, i.e.
\be
\bar{\r}(t_0) = \r(t_0)\quad \Longleftrightarrow \quad [\r(t_0), H(t_0)]=0.
\ee{rbr}

In particular, if the system is initially in a canonical state the
known form of the canonical characteristic function results
\cite{TLH}. From this very result the Tasaki-Crooks fluctuation theorem and the
Jarzynski work theorem follow readily \cite{TLH,TH}. Notably, the
characteristic function (\ref{Gu}) 
assumes the form of a two-point quantum
correlation as it is the case in linear response theory, despite the fact that 
it embraces the full nonlinear response on the perturbation by the
applied force.

\section{Fluctuation theorems for microcanonical initial states}
\label{III}
A microcanonical initial state is characterized by an ensemble that
consists of all eigenstates of $H(t_0)$ with eigenvalues in a
narrow band around an energy $E$. All contributing states have equal
weights. In the ideal case of an arbitrarily narrow energy range around
$E$ the state of the system is given by the following density matrix
\be
\begin{split}
\r_{E}(t_{0})& = \o^{-1}_{E}(t_{0})\: \d(H(t_{0})-E)\\
&=\o^{-1}_{E}(t_{0})\: \sum_{k} \d\big (e_{k}(t_{0} )-E \big ) \:P_{k}(t_{0}) ,
\end{split}
\ee{rmc}
where 
\be 
\begin{split}
\o_{E}(t_{0}) &= \Tr \:\d(H(t_{0})-E)\\
&= \sum_{k} g_{k}\:\d\big (e_{k}(t_{0})-E \big)
\end{split}
\ee{o} 
denotes the density of states of the system
described by the Hamiltonian $H(t_{0})$, and $g_{k} = \Tr \:P_{k}(t_{0})$ the degeneracy of
the energy $e_{k}(t_{0})$. 
Eq. (\ref{rmc}) presents a 
rather formal expression for the density matrix. It can be
regularized in the standard way by replacing the delta function by any
smooth function $\d_{\e}(H(t_{0})-E)$ which approaches the delta
function in the limit $\e \to 0$. The value of $\e$ may be chosen
such that the width of $\d_{\e}(H(t_{0})-E)$ is of the order of the
accepted uncertainty of energy. It is also convenient to take an
approximating delta function with infinite support such as a Gaussian
function in order to have a well defined density matrix for all
energies $E$ even if $E$ falls in a gap of the spectrum. Since the
width $\e$ is always assumed to be much smaller than all other
relevant energies in all following expressions the formal
manipulations with delta functions also apply for the respective
expressions containing the regularized delta functions.
Combining eqs.(\ref{Gu}) and (\ref{rmc}) we find the following
expression for the
characteristic function of work 
\begin{align}
G_{t_{f},t_{0}}(E,u) =&\: \o^{-1}_{E}(t_{0})\:\nonumber \\ 
&\:\times \Tr\: e^{iuH_{H}(t_{f})}\:
e^{-iuH_{H}(t_{0})}\: \delta(E-H(t_{0})) \nonumber \\
=&\: \o^{-1}_{E}(t_{0}) \nonumber \\
&\:\times \Tr \:e^{iuH_{H}(t_{f})-E}\:
\delta(H(t_{0})-E).  
\label{Gumc}
\end{align}
The inverse Fourier transform is readily performed to yield the
probability density of work,
\begin{align}
p_{t_{f},t_{0}}(E,w) =&\: \int \frac{du}{2 \p} e^{-iuw}\:
G_{t_{f},t_{o}}(E,u) \nonumber \\
=&\: \o^{-1}_{E}(t_{0})\nonumber \\
&\:\times \Tr \:\d (H_{H}(t_{f})-E-w) \:\d(H(t_{0})-E). 
\label{pmc}
\end{align}
From this expression one may formally obtain the corresponding
classical result by replacing the Hamilton operators by the
corresponding Hamilton functions and the trace by an integral over all
possible initial states. Thereby, the Hamiltonian at the
final time, $H_{H}(t_{f})$ must be replaced by the Hamilton function 
depending on the
initial phase space variables $\bz_{0}$ via the solutions $\bz(\bz_{0},t_{f})$ 
of Hamilton's
equations of motion at the final time. This
results in the expression
\be
\begin{split}
p^{\text{cl}}_{t_{f},t_{0}}(E,w) =&\: \o^{-1}_{E}(t_{0})\\
&\:\times \i d \bz_{0}\:
\d(H(\bz(\bz_{0},t_{f}),t_{f})-E-w)\\ 
&\:\times \d(H(\bz_{0},t_{0})-E)
\end{split}
\ee{pcl} 
which agrees with the result obtained on the basis of classical
statistical mechanics in Ref.~\cite{CvdBK}. 
In the classical formulation it is of course correct to replace the
energy $E$ in the first delta function by the Hamilton function at the
initial time as dictated by the second delta function. This
transformation is not possible in the quantum 
expression confirming  the recent observation that work is not a quantum
mechanical observable \cite{TLH}.

\subsection{Microcanonical quantum Crooks theorem}
The dependence on the work $w$  can be shifted in
eq. (\ref{pmc})
from the first to the second delta
function  by introducing the final
energy $E_{f}=E+w$. For the product of the density of states
$\o_{E}(t_{0})$ and the
probability density $p_{t_{f}t_{0}}(E,w)$ one obtains
\begin{align}
\o_{E}(t_{0}) p_{t_{f},t_{0}}(E,w) =&\: \Tr\: \d\left ( H_{H}(t_{f})-
  E_{f}\right ) \nonumber \\
&\:\times \d \left (H(t_{0}-E_{f}+w) \right) \nonumber \\
=&\:\Tr \: \d \left ( H(t_{f})-E_{f}\right)\nonumber \\
&\:\times \d \left (
  \bar{H}_{H}(t_{0})-E_{f} +w\right) \nonumber \\
&= \o_{E+w}(t_{f})\: p_{t_{0},t_{f}}(E+w,-w), 
\label{qC}
\end{align}
where, in going to the second line, we used the cyclic invariance of the
trace and the unitarity of the time evolution in order to transfer the time
dependence from  the first to the second delta function. Here the
operator $\bar{H}_{H}(t_{0})$ denotes the Hamiltonian in the
Heisenberg picture for the time evolution running in reversed order
with a backward protocol from the
final to the initial time, i.e.:
\be
\bar{H}_{H}(t_{0}) = U^{+}_{t_{0},t_{f}} H(t_{f}) U_{t_{0},t_{f}}
\ee{Hb}
In the third line of eq. (\ref{qC}) the result was again written
in terms of the probability density of work as a function of the initial
energy $E$. This relation can be
formulated 
as the microcanonical quantum version of Crooks' theorem \cite{C,TH,CvdBK}
saying that
\be
\frac{p_{t_{f},t_{0}}(E,w)}{p_{t_{0},t_{f}}(E+w,-w)}= e^{\left
    [S_{\o}(E+w,t_{f})-S_{\o}(E,t_{0}) \right ]/k_{B}}.
\ee{mcC}
It relates the ratio between the probability densities of work for a process 
and the corresponding time reversed process to the difference of 
the entropies which belong to microcanonical equilibrium
systems with Hamiltonians $H(t_{0})$ and $H(t_{f})$. The entropies are
defined in terms of the respective density of states via the standard
statistical mechanical relation
\be
S_{\o}(E,t) = k_{B}\ln \o_{E}(t),
\ee{So} 
where $k_{B}$ denotes the Boltzmann constant. Note that the
  similarly looking canonical Crooks theorem \cite{C} contains a
  nonequilibrium entropy production $\o_{\text{Crooks}}$ which
  is given by
  $T \o_{\text{Crooks}}=T S_{\text{Crooks}}
  -Q= w-\D F $ where  $S_{\text{Crooks}}$ denotes  the nonequilibrium entropy,
  $Q$ the heat, $\D F$ the thermodynamic free energy  and $w$ the
  nonequilibrium 
  work. This entropy production is distinctly different from the
  statistical mechanical expression (\ref{So}) for the
  microcanonical entropy entering eq. (\ref{mcC}).  

Actually, a yet different statistical mechanical entropy can be
  defined. It is proportional to the logarithm of 
the number of states
below the energy $E$ rather than of the density of states at the
energy $E$ \cite{H}. This entropy then takes the form
\be
S_{\O}(E,t) = k_{B}\ln \O_{E}(t) 
\ee{SO}
where 
\be
\O(E,t) = \i_{-\infty}^{E} dE' \o(E',t)
\ee{O}
denotes the number of states below $E$.
For systems with short ranged interactions the two definitions are
known to coincide in the thermodynamic limit, i.e. in the limit of
infinite systems. For small systems though  
$S_{\O}(E)$  has been proved to be more advantageous as it is an increasing
function of energy by definition.
\cite{PHT,entropy}. 

We note that for an ensemble of initial states with energies uniformly
  distributed up to the energy $E$,  
a fluctuation theorem of the form of eq. (\ref{mcC}) can be
derived, in which the difference 
of the entropies $S_{\o}$ is replaced by the corresponding difference
of $S_{\O}$, i.e.
\be
\frac{p^{<}_{t_{f},t_{0}}(E,w)}{p^{<}_{t_{0},t_{f}}(E+w,-w)}= e^{\left
    [S_{\O}(E+w,t_{f})-S_{\O}(E,t_{0}) \right ]/k_{B}}.
\ee{mcCO}
This follows along the same line of arguments leading to the
  relation (\ref{mcC}) 
for
the uniform initial density matrix $\r^{<}(t_{o})$ which is given by
\be
\r_{<E}(t_{0}) = \O_{E}(t_{0}) \Th (E-H(t_{0}))
\ee{rmE}
and which was introduced by Ruelle \cite{R} in the context of the
microcanonical ensemble. The probability
of work $p^{<}_{t_{f},t_{0}}(E,w)$ for this ensemble 
is obtained from the corresponding
microcanonical probability by an integration over the energy, i.e.
\be
p^{<}_{t_{f},t_{0}}(E,w)= \O^{-1}(E,t_{0}) \i_{-\infty}^{E} dE'\:\o(E',t_{0})
p_{t_{f},t_{0}}(E',w).    
\ee{plp}

The fact that the microcanonical quantum Crooks theorem (\ref{mcC}) 
depends on the
time reversed process seemingly restricts its practical usefulness. As
opposed to
computer experiments it is impossible to perform an active
reversal of time, i.e. to let time run backwards,
in real experiments. 

\subsection{Entropy-from-work theorem}
The experimentally inaccessible probability density of the time
reversed process  
though can be eliminated by first expressing the initial energy in terms of
the final energy and the performed work, and next, by integrating
eq. (\ref{qC}) over all possible values of the work. In this way, the
density of states  at the later time can be expressed by an
integral of the initial density of states, 
weighted by $p_{t_{f},t_{0}}(E_{f}-w,w)$, i.e. 
\begin{align}
\i dw\: \o_{E_{f}-w}(t_{0}) &p_{t_{f},t_{0}}(E_{f}-w,w)
\nonumber \\
=&\:
\i dw\: \o_{E_{f}}(t_{f}) \:p_{t_{0},t_{f}(E_{f},-w)} \nonumber\\
=&\: \o_{E_{f}}(t_{f}) 
\label{EWR}
\end{align}
With the definition (\ref{So}) the following relation
  between the entropy 
of the initial system and the unknown entropy of the final system can
be established:
\be
\i dw \:e^{S_{\o}(E_{f-w}(t_{0}))} \:p_{t_{f},t_{0}}(E_{f}-w,w) =
e^{S_{\o}(E_{f})(t_{f})}. 
\ee{EWR2}

Note, that for a fixed final energy the weighting function 
$p_{t_{f},t_{0}}(E_{f}-w,w)$ is not a probability density of the
performed work. In the following we demonstrate that 
the left hand side of eq.~(\ref{EWR}) can be written in terms of a
properly defined 
average of the exponentiated entropy conditioned on the final energy.

In general, from a single initial energy not all relevant final
energies are likely to be reached, or may even be impossible to
reach. 
Therefore, in an experiment the initial energies
have to be scanned over a sufficiently large range of values. For each initial
energy a sufficient number of experiments has to be performed in order that
a reliable statistics of work can be compiled for conveniently binned
final energies. Based on such a statistics the probability density of
initial energies $E$ conditioned on the final energy $E_{f}$, 
$\r_{t_{f},t_{0}}(E|E_{f})$, can be inferred. If the initial energies
$E$ are uniformly sampled in a range of energies of size $E_{R}$ the
joint probability of initial and final energies is given by
$\r(E,E_{f}) = p_{t_{f},t_{0}}(E,E_{f}-E)/E_{R}$. According to Bayes
theorem the conditional
probability density $\r_{t_{f},t_{0}}(E|E_{f})$ becomes
\begin{align}
\r_{t_{f},t_{0}}(E|E_{f}) &= \frac{\r(E,E_{f})}{\i dE \r(E,E_{f})}
\nonumber \\
&= \frac{p_{t_{f},t_{0}}(E,E_{f}-E)}{\i dE
  \:p_{t_{f},t_{0}}(E,E_{f}-E) }.  
\label{rp}
\end{align}
Consequently, 
the integral over the density of
  states
can be formulated as an
average over initial energies conditioned on the final energy, yielding
\begin{align}
\langle \o_{E}(t_{0}) \rangle_{E_{f}} \equiv&\: \i dE \:
\o_{E}(t_{0}) \r_{t_{f},t_{0}}(E|E_{f}) \nonumber \\
=&\: N(E_{f})^{-1} \nonumber \\
& \times \i dw \: \o_{E_{f}-w}(t_{0})
p_{t_{f},t_{0}}(E_{f}-w,w)
\label{eSE} 
\end{align}
where
\be
N(E_{f}) = \i dE
  \:p_{t_{f},t_{0}}(E,E_{f}-E)
\ee{N}
guarantees the normalization of the conditional average.
Using eq. (\ref{EWR}) one finds the density of states
$\o_{E_{f}}(t_{f})$ of a microcanonical  
system with the Hamiltonian $H(t_{f})$ represented by
the  average of the density of states  $\o_{E}(t_{0})$ with respect the
conditional probability $\r_{t_{f},t_{0}}(E|E_{f})$ reading 
\be
\o_{E_{f}}(t_{f}) = N(E_{f})  \langle
\o_{E}(t_{0})\rangle_{E_{f}}. 
\ee{dsfds}
Expressing the density of states in terms of the entropy (\ref{So})
one obtains
\be
e^{S_{\o}(E_{f},t_{f})/k_{B}} = N(E_{f})  \langle
e^{S_{\o}(E)/k_{B}}\rangle_{E_{f}}. 
\ee{SfSo}
We call 
eq. (\ref{SfSo}) the
''entropy-from-work'' theorem.       
In analogy to the Jarzynski relation it allows one to extract
equilibrium properties from
non-equilibrium experiments, which are, in the case of the Jarzynski
relation, the free energy and, in the present context, 
the entropy $S_{\o}(E)$.

Based on the relation (\ref{mcCO}) 
an entropy-from-work theorem for 
$S_{\O}(E)$
follows in close analogy to
the corresponding relations
(\ref{dsfds}) and 
(\ref{SfSo}) reading
\be
\O_{E_{f}}(t_{f}) = N^{<}(E_{f})  \langle
\O_{E}(t_{0})\rangle^{<}_{E_{f}}. 
\ee{NsNs}
and
\be
e^{S_{\O}(E_{f},t_{f})/k_{B}} = N^{<}(E_{f})  \langle
e^{S_{\O}(E)}\rangle^{<}_{E_{f}/k_{B}}. 
\ee{SfSO}
where the average $\langle \cdot \rangle^{<}_{E_{f}}$ is performed
with respect to the conditional probability
density
\be
\r^{<}_{t_{f},t_{0}}(E|E_{f}) =
 N^{<}(E_{f})^{-1}\: p^{<}_{t_{f},t_{0}}(E,E_{f}-E).  
\ee{rfk}
and where
\be
N^{<}(E_{f})=
\i dE
  \:p^{<}_{t_{f},t_{0}}(E,E_{f}-E).
\ee{Nk}

\subsection{Interrelations with the characteristic function of work
  for canonical initial states}
To establish a relation between the microcanonical and
canonical work distributions we integrate the  expression
(\ref{Gumc}) for the 
micocanonical characteristic function of work  
weighted by the density of states and by a Boltzmann factor $\exp(-\b
E)$ with inverse temperature $\b$.  Performing this integration under
the trace we obtain
\begin{align}
\i dE \:\o_{E}(t_{0})\: e^{-\beta E} &G_{t_{f},t_{0}}(E,u) \nonumber
\\
& = \Tr \:e^{iu
  H_{H}(t_{f})}\: e^{-iu H(t_{0})}\: e^{-\b H(t_{0})} \nonumber \\
&= Z(t_{0}) \:G^{\b}_{t_{f},t_{0}}(u)
\label{Gmcc}
\end{align}
where 
\be
 G^{\b}_{t_{f},t_{0}}(u) = Z_{\b}(t_{0})^{-1} \Tr \:e^{iuH_{H}(t_{f})}
 e^{-iuH(t_{0})} e^{-\b H(t_{0})}
\ee{Gc}
denotes the characteristic function of work for a process starting from
a canonical density matrix \cite{TLH} and 
\be
Z_{\b}(t_{0}) = \Tr \:e^{-\b H(t_{0})}
\ee{Z}
the respective partition function.
Hence, the canonical characteristic function of work is
related to the microcanonical one by a Laplace transform. 
The reverse relation is readily obtained by the following inverse Lapace
transform 
\be
 G_{t_{f},t_{0}}(E,u) =\o_{E}(t_{0})^{-1}
\i_{{\mathcal C}}\frac{d \b}{2 \p i}\: Z_{\b}(t_{0})\:e^{\b E}\:
G^{\b}_{t_{f},t_{0}}(u)   
\ee{Gcmc}
where ${\mathcal C}$ is an inverse Laplace contour in the complex $\b$
plane from $-i
\infty+c$ to $i \infty+c$. The real constant $c$ must be chosen such
that all singularities of the integrand lie to its
left side. The density of states is related to the partion function by
the standard relation
\be
\o_{E}(t_{0}) = \i_{{\mathcal C}} \frac{d \b}{2 \p i} \:
Z_{\b}(t_{0})\: e^{\b E}.
\ee{oZ}

Accordingly, the probability densities of work for microcanonical and
canonical initial states are also related by a Laplace transform,
yielding
\begin{align}
p^{\b}_{t_{f},t_{0}}(w) &= Z_{\b}(t_{0})^{-1} \i dE\:\o_{E}(t_{0})\:
e^{-\b E}\:p_{t_{f},t_{0}}(E,w)\\  
p_{t_{f},t_{0}}(E,w)&= \o_{E}(t_{0})^{-1} \i_{\mathcal C} \frac{d
  \b}{2 \p i}\: Z_{\b}(t_{0})\: e^{\b E} \:p^{\b}_{t_{f},t_{0}}(w)
\end{align}  
Finally we note that the Jarzinski relation can be obtained from the
microcanonical Crooks theorem. For this purpose one considers the left
hand side of the first line and the right hand side of the third line
of eq. (\ref{qC}), multiplies both sides with $\exp \left (- \b (E+w)
\right)$ and integrates over all values of $E$ and $w$. For the left
hand side one then finds:
\begin{align}
\i dE \i dw \:e^{-\b(E+w)}&\o_{E}(t_{0}) \:p_{t_{f},t_{0}}(E,w)
\nonumber\\ 
=&\: \i dw\:
e^{-\b w} \:Z_{\b}(t_{0}) \:p^{\b}_{t_{f},t_{0}}(w) \nonumber \\
=&\: Z_{\b}(t_{0}) \:\langle e^{-\b w} \rangle
\end{align}
while the right hand side yields
\begin{align}
\i dE \i dw \:e^{-\b(E+w)} &\o_{E+w}(t_{f})
\:p_{t_{0},t_{f}}(E+w,-w)\:\nonumber \\
=&\:
\i dE_{f} \:e^{-\b(E_{f})} \:\o_{E_{f}}(t_{f}) \nonumber \\
&\:\times \i dw \:p_{t_{0},t_{f}}(E_{f},-w) \nonumber \\
=&\: Z_{\b}(t_{f})
\end{align}
where we substituted the integration variable $E$ by $E_{f} =E+w$.
A comparison of the last two equations immediatly yields the
Jarzynski relation \cite{J,TLH}.         

\section{Conclusions}
The expression for the characteristic function of work performed on an 
isolated quantum system by an external force was generalized for
arbitrary initial states. The general structure of the
characteristic function is given by a correlation function of the
exponentiated system Hamiltonians at the first and the second
measurement time. The quantum expectation and ensemble
average are jointly taken with
respect to a density matrix which results from the actual initial state
immediately  before the first measurement by means of a state reduction
with respect to the
energy eigenbasis of the then measured Hamiltonian. 
Initial states which commute with the
Hamiltonian at the initial time consequently are not modified.
This form of the characteristic function holds irrespectively of a
possible degeneracy of the spectrum of the system's Hamiltonian.

For a microcanonical initial state expressions for the
characteristic function of work and the respective probability density
were established. In the classical limit known expressions  were recovered 
and the validity of the microcanonical Crooks
fluctuation theorem was demonstrated for quantum systems.
Moreover we formulated an entropy-from-work theorem which allows one
to infer the unknown entropy of a system from a reference system with known
entropy by means of a nonequilibrium experiment.  In such an
experiment the initial system with known entropy is deformed into the
final system within finite time according to a prescibed protocol of
Hamiltonians. In this context we want to emphasize that the
  entropy following from this theorem is based on either of the {\it
 statistical 
  mechanical} definitions (\ref{So}), or (\ref{SO})  which
agree with each other for sufficiently large systems with short range
interactions. In
  contrast, the canonical Crooks theorem contains a nonequilibrium
  entropy production that 
  emerges from a relation involving free energy, heat and
  nonequilibrium work.

We further note that the microcanonical distribution 
provides the appropriate description of an {\it isolated} system
if no further information about its state is available 
even if the energy of the considered system is a priori unknown. 
By registering the result of the first energy measurement this ignorance is
removed and the available information about  the initial state is expressed
without any bias by the microcanonical density matrix corresponding to
the measured energy.  
Though, if the unforced dynamics does not only leave invariant the energy
but also other quantities such as linear or angular total momentum,
then the adequate constrained microcanonical ensemble has to be
considered as the 
proper initial state.

Finally, by means of properly weighted Laplace transforms, relations
between microcanonical and canonical characteristic functions and
probability densities of work were established.

{\bf Acknowledgment.}
The authors thank Prof. K. Sch\"onhammer for valuable comments on
the manuscript and
J. Dunkel for many very helpful discussions.
This work was supported by the SFB 438, project A10, the
Nanosystems Initiative Munich (NIM)(PH,PT) as well as by the Ministerio de
Educaci\'on y Ciencia of Spain (FIS2005-02884) and the Junta de Andalucia
(MM).

\end{document}